\documentclass[aps,prb,twocolumn, floatfix]{revtex4-1}
\usepackage{graphics,bm}
\usepackage{amssymb}
\usepackage{epsfig}
\usepackage{epsf}
\usepackage[usenames]{color}
\usepackage{upgreek}
\usepackage{amsmath}
\usepackage{xcolor}
\usepackage{hyperref}
\setcitestyle{numbers,square}

\renewcommand{\vec}[1]{\bm{ {\bf #1}}}

\begin{document}

\title{Temperature dependence of the magnon spin diffusion length and magnon spin conductivity in the magnetic insulator yttrium iron garnet}
\author{L.J. Cornelissen}
\email{l.j.cornelissen@rug.nl}
\affiliation{
    Physics of Nanodevices, Zernike Institute for Advanced Materials,
    University of Groningen,
    Nijenborgh 4,
    9747 AG Groningen,
    The Netherlands
}

\author{B.J. van Wees}
\affiliation{
    Physics of Nanodevices, Zernike Institute for Advanced Materials,
    University of Groningen,
    Nijenborgh 4,
    9747 AG Groningen,
    The Netherlands
}

\begin{abstract}
We present a systematic study of the temperature dependence of diffusive magnon spin transport, using a non-local device geometry. In our measurements, we detect spin signals arising from electrical and thermal magnon generation, and we directly extract the magnon spin diffusion length $\lambda_m$ for temperatures from 2 to 293~K. Values of $\lambda_m$ obtained from electrical and thermal generation agree within the experimental error, with $\lambda_m=9.6\pm0.9$~$\upmu$m at room temperature to a minimum of $\lambda_m=5.5\pm0.7$~$\upmu$m at 30~K. Using a 2D finite element model to fit the data obtained for electrical magnon generation we extract the magnon spin conductivity $\sigma_m$ as a function of temperature, which is reduced from $\sigma_m=5.1\pm0.2\times10^5$~S/m at room temperature to $\sigma_m=0.7\pm0.4\times10^5$~S/m at 5 K. Finally, we observe an enhancement of the signal originating from thermally generated magnons for low temperatures, where a maximum is observed around $T=7$~K. An explanation for this low temperature enhancement is however still missing and requires additional investigations.
\end{abstract}
\maketitle

Magnons, the quanta of spin waves, can be excited in magnetic insulators in various ways: magnetically via microwave-frequency AC currents \cite{Chumak2015}, thermally via the spin Seebeck effect (SSE) \cite{Uchida2010} or electrically via low-frequency or DC electric currents making use of the interfacial spin-flip scattering mechanism. The latter excitation method has attracted a lot of attention recently, both experimentally \cite{Cornelissen2015, Goennenwein2015, Velez2016, Li2016, Wu2016, Cornelissen2016} and theoretically \cite{Xiao2015,Zhang2012,PhysRevLett.109.096603, Cornelissen2016a}. It relies on the exchange coupling between the spin accumulation in a normal metal (NM) and magnons in a magnetic insulator (MI), where the materials of choice are typically platinum (Pt) for the NM and yttrium iron garnet (YIG) as the MI. Via this exchange coupling, spin current can be transferred between the MI and the NM. The spins in the MI are then carried by magnons and transported diffusively, allowing for the definition of a magnon spin diffusion length ($\lambda_m$) and a magnon spin conductivity ($\sigma_m$) analogous to their counterparts in diffusive electron spin transport \cite{Cornelissen2016a}. Several recent experiments investigated the temperature dependence of diffusive magnon spin currents \cite{Goennenwein2015, Velez2016, Li2016, Wu2016}, however no systematic study of $\lambda_m$ and $\sigma_m$ as a function of temperature has been carried out to date. Recently, the relevant length scale for the local SSE was measured as a function of temperature \cite{Guo}, which as we show here exhibits a different temperature dependence than $\lambda_m$. Additionally, Giles \emph{et al.} extracted $\lambda_m$ using experiments where magnons are generated via laser heating and found $47\leq\lambda_m\leq73$ $\upmu$m at 23 K \cite{Giles2015a}, and an upper bound of $\lambda_m\leq10$ $\upmu$m at 280 K. However, they did not report the full temperature dependence of $\lambda_m$, which we do identify here.

\begin{figure*}
	\includegraphics[width=15.0cm]{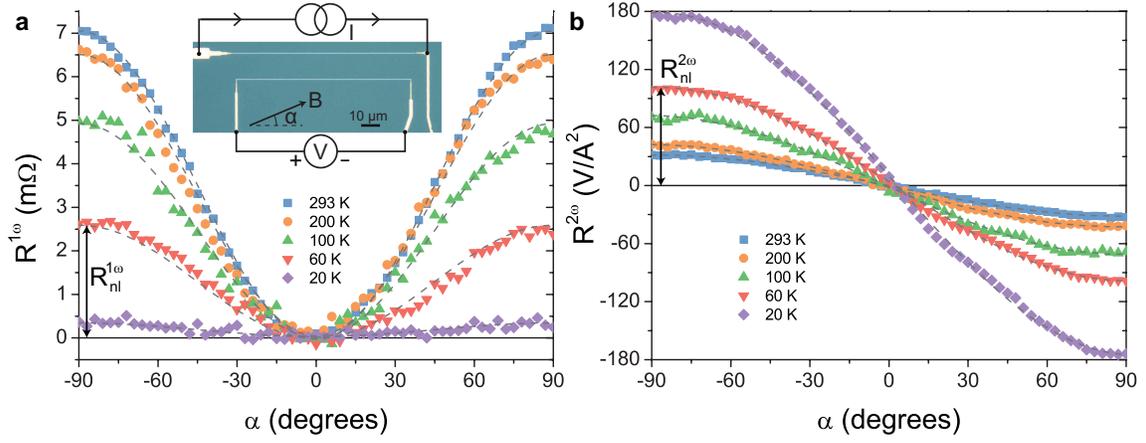}
	\caption{Non-local signals as a function of the angle $\alpha$ between the magnetic field $B$ and the injector/detector strips, for injector-detector separation distance $d=3.5$ $\upmu$m and various sample temperatures. Inset shows an optical microscope image of one of the devices, with current and voltage connections indicated schematically. a) First harmonic signal. Dashed lines are $\sin^2(\alpha)$ fits through the data. b) Second harmonic signal. Dashed lines are $\sin(\alpha)$ fits through the data. The amplitudes of the non-local signal, $R_{\rm nl}^{1\omega}$ and $R_{\rm nl}^{2\omega}$ are indicated in figure a and b respectively, for $T = 60$ K. The sign convention is the same as in Ref.~\cite{Cornelissen2015}, meaning that a positive $R_{\rm nl}^{2\omega}$ implies a second harmonic voltage that is \emph{opposite} to what would be obtained for a local current driven spin Seebeck measurement.
	}
	\label{fig:angle_sweeps}
\end{figure*}

\begin{figure*}
	\includegraphics[width=15.0cm]{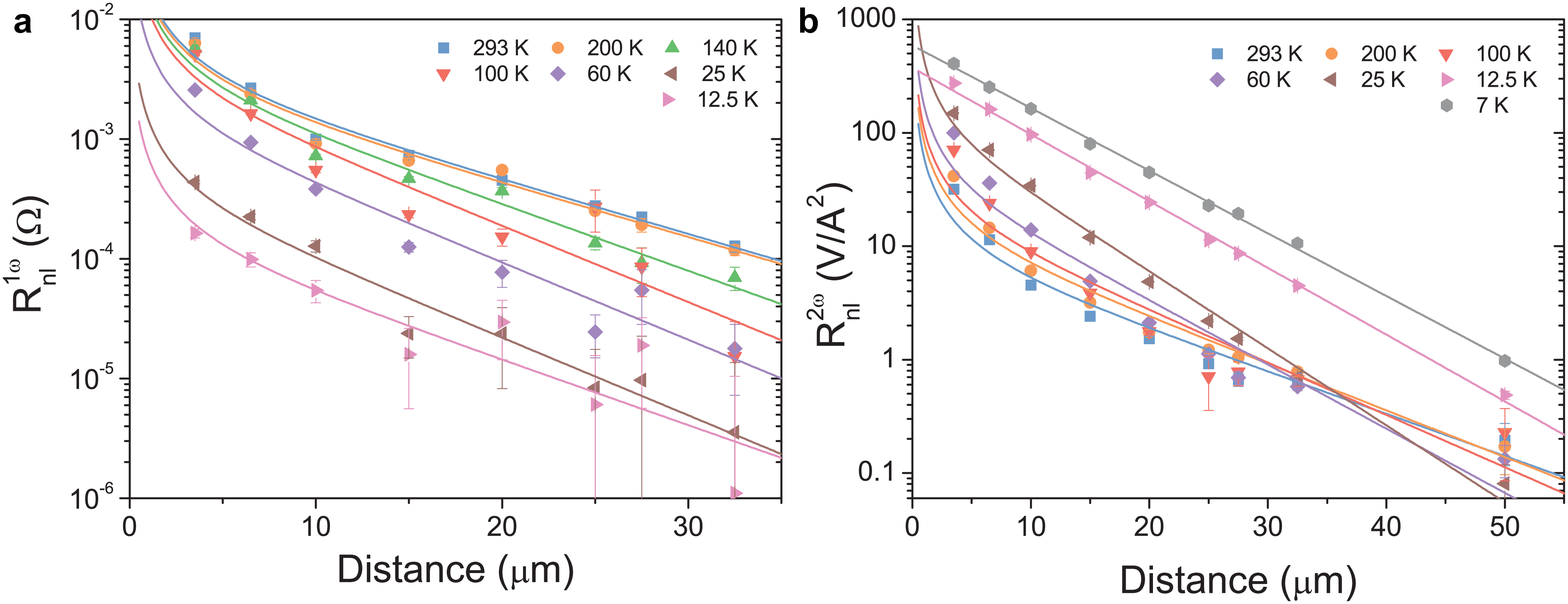}
	\caption{Distance dependence of the amplitudes of the non-local first (a) and second (b) harmonic signals. Errorbars indicate one standard error of the amplitude obtained from the fits to the angle dependent data. Solid lines are fits of the data to Eq.~\ref{eq:mspt}. Exception to this are the solid lines in the low temperature ($T\leq12.5$ K) second harmonic data, which are better described by a pure exponential fit.
	}
	\label{fig:dist_dep}
\end{figure*}

In this paper we investigate the diffusive transport of magnon spins as a function of sample temperature. We employ the non-local measurement geometry that was developed in our earlier work \cite{Cornelissen2015} in which we measure the magnon spin signal as a function of distance, which allows us to directly extract $\lambda_m$ for temperatures from 2~K to 293~K. In this measurement scheme, magnon injection and detection results from the exchange interaction between a spin accumulation in the platinum injector and detector (created and probed by the spin Hall and inverse spin Hall effect, respectively) and magnons in the YIG. This implies that the distance over which the magnon spin current diffuses is well defined since the locations of both magnon injection and detection are strictly determined, allowing us to unambiguously extract the magnon spin diffusion length \cite{Cornelissen2016}. Additionally, we use a 2D finite element model (FEM) to describe the magnon transport in our devices \cite{Cornelissen2016a}, which enables us to determine $\sigma_m$ as a function of temperature.

A microscope image of a typical device is shown in the inset of Fig.~\ref{fig:angle_sweeps}. The devices consist of two parallel platinum strips on top of a yttrium iron garnet (YIG) thin film, separated a distance $d$ from each other and contacted by Ti/Au leads. The YIG film is 210~nm thin and was grown by liquid phase epitaxy in the (111) direction, on top of a 500~$\upmu m$ thick Gd$_3$Ga$_5$O$_{12}$ substrate. YIG\textbar{}GGG samples were obtained commercially from Matesy GmbH. Three steps of electron beam lithography were used to define the devices on top of the YIG film. In the first step we define a pattern of Ti/Au markers (deposited by e-beam evaporation), used to align the subsequent steps. Injectors and detectors are defined in the second step, where approximately 10~nm of platinum is deposited using magnetron sputtering in an Ar$^+$ plasma. Finally, we define Ti/Au (5/75~nm) leads and bonding pads using e-beam evaporation. Prior to Ti evaporation, we perform argon ion milling to remove any polymer residues from the Pt strips. Length and width of the platinum strips are approximately $L=100$~$\upmu$m and $w=300$~nm for all devices. 

Non-local measurements are performed by applying an AC charge current $I$ to the injector  at a frequency $\omega$ (typically $I_{\rm rms}=100$~$\upmu$A and $\omega/(2\pi)=3.423$~Hz). This current generates magnons in the YIG via two different mechanisms: Due to the spin Hall effect, a transverse spin current is generated towards the YIG and a spin accumulation $\mu_s$ builds up at the Pt\textbar{}YIG interface. Via the exchange interaction at the interface, $\mu_s$ generates a magnon spin accumulation $\mu_m$ in the YIG. This is a fully linear process, i.e. $\mu_m\propto I$. Additionally, heat is generated in the injector via Joule heating, which induces a temperature gradient $\nabla T$ in the YIG. By virtue of the spin Seebeck effect, this gradient causes a magnon spin current to flow. The spin current $\vec{j}_m$ is linear with the temperature gradient, which in turn is proportional to the current \emph{squared}, i.e. $\vec{j}_m\propto\nabla T\propto I^2$. At the detector interface, magnon spins in the YIG are converted into a spin accumulation in the Pt, which is then converted to a charge voltage $V$ via the inverse spin Hall effect. Using a lock-in detection technique\cite{Vlietstra2014} we can detect signals due to processes that are linear and quadratic in the current separately. The non-local first harmonic signal is then given by $R^{1\omega}=V^{1\omega}/I$ (due to electrical generation), while the second harmonic signal is $R^{2\omega}=\sqrt{2}V^{2\omega}/I^2$ (thermal generation). In the non-local measurements we are thus sensitive to the generation, transport and detection of magnons, where the only difference between first and second harmonic lies in the generation process.

We now rotate the sample in an external in plane magnetic field large enough to align the YIG magnetization $\vec{M}$ ($B=10$~mT), thus varying the angle $\alpha$ between $\vec{M}$ and the Pt strips. For electrical generation of magnons, both the injection and detection processes depend on the projection of $\vec{M}$ on the spin accumulation in the Pt, which leads to $R^{1\omega}=R_{\rm nl}^{1\omega}\sin^2(\alpha)$ for the first harmonic signal as can be seen in Fig.~\ref{fig:angle_sweeps}a. For thermally generated magnons, only the detection depends on $\alpha$, resulting in $R^{2\omega}=R_{\rm nl}^{2\omega}\sin(\alpha)$ for the second harmonic signal, as can be observed in Fig.~\ref{fig:angle_sweeps}b. The first harmonic signal decreases for decreasing sample temperature, which is consistent with previous observations \cite{Goennenwein2015, Li2016,Velez2016} and theoretical predictions \cite{bender2012,Zhang2012,Cornelissen2016a}. Interestingly, the second harmonic signal shows the opposite trend and significantly increases as $T$ is reduced. 

By performing measurements for various injector-detector separation distances $d$, we extract the signal amplitude as a function of distance. The results are shown in Fig.~\ref{fig:dist_dep}a for the first harmonic signal and Fig.~\ref{fig:dist_dep}b for the second harmonic signal, for several temperatures. For large distances, the decay of the magnon spin signal is governed by the magnon spin diffusion length $\lambda_m$. As we showed in Ref.~\cite{Cornelissen2015}, $\lambda_m$ can be extracted from both the first and second harmonic signals by fitting the distance dependent data \footnote{\label{fn:fit}To ensure that data from all distances is weighed equally, we perform the fit on the $\log_{10}$ of the data, using also the $\log_{10}$ of Eq.~\ref{eq:mspt} as fitting function.} to the 1D magnon spin diffusion model
\begin{equation}
\label{eq:mspt}
R_{\rm nl}(d)=\frac{A}{\lambda_m}\frac{\exp(d/\lambda_m)}{1-\exp(2d/\lambda_m)} ,
\end{equation}
where $A$ is a prefactor that depends for instance on the efficiency of the magnon injection (governed by the effective spin conductance $g_s$ \cite{Cornelissen2016a}) and on the magnon diffusion constant. The model in Eq.~\ref{eq:mspt} assumes transparent injector and detector contacts, a condition that even at room temperature is not completely fulfilled due to the finite value of $g_s$. Additionally, $g_s$ scales as $g_s\propto\left(\frac{T}{T_C}\right)^{3/2}$ \cite{bender2012,Cornelissen2016a} so that for $T\rightarrow0$ the interfaces become increasingly opaque, making the applicability of the model questionable for low temperature. However, we can still use the model to determine $\lambda_m$, since that is only determined by the decay of the signal in the long-distance regime (i.e. the slope of the curves in Fig.~\ref{fig:dist_dep} for distances $d>10$~$\upmu$m). For $T<25$~K the signal-to-noise ratio (SNR) in the first harmonic is~$\ll1$ for devices with $d>10$~$\upmu$m, such that we can no longer reliably extract $\lambda_m$. 

The distance dependence of the second harmonic is generally more complicated than that of the first harmonic due to the delocalized nature of thermal magnon generation, even showing a sign change for very short distances ($d\leq200$~nm) as we showed in Ref.~\cite{Cornelissen2015}. However, for longer distances we can still use the model in Eq.~\ref{eq:mspt} to extract $\lambda_m$ from the second harmonic data. Interestingly, for $T<25$~K the second harmonic distance dependence shows almost pure exponential decay described by
\begin{equation}
\label{eq:exp}
R_{\rm nl}^{2\omega}=B\exp(-d/\lambda_m^{2\omega})\,,
\end{equation}
over approximately three orders of magnitude. In this regime we thus extract $\lambda_m^{2\omega}$ by fitting the data to Eq.~\ref{eq:exp}. This crossover to a pure exponential might be explained by the reduction of interface transparency due to the decrease of $g_s$. This can also be observed in spin transport in metallic non-local spin valves, where transparent contacts result in signal decay similar to our Eq.~\ref{eq:mspt}, but opaque contacts yield pure exponential decay \cite{Takahashi2003}. 
\begin{figure}
	\includegraphics[width=8.0cm]{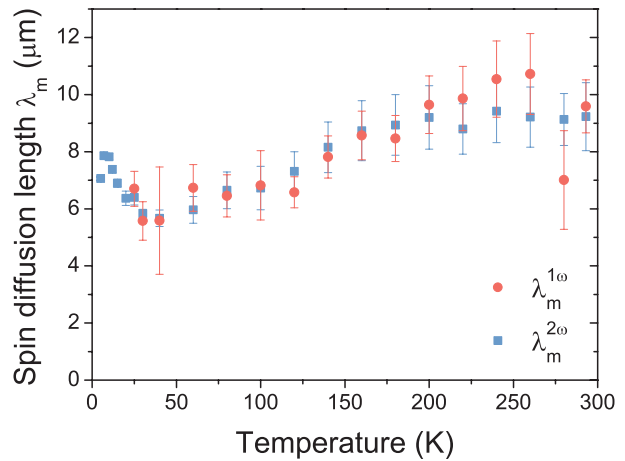}
	\caption{Magnon spin diffusion length as a function of temperature, obtained from the distance dependence of the first harmonic ($\lambda_m^{1\omega}$) and second harmonic ($\lambda_m^{2\omega}$) signals. $\lambda_m$ was extracted by fitting the data to Eq.~\ref{eq:mspt}. Errorbars indicate one standard error obtained from the fits. For $T\geq25$~K, $\lambda_m^{1\omega}$ and $\lambda_m^{2\omega}$ agree within the experimental uncertainties. For $T <25$~K, the signal-to-noise ratio (SNR) in the first harmonic is~$\ll$~1 for distances $d>10$~$\upmu$m, making reliable extraction of $\lambda_m^{1\omega}$ for impossible. However, due to the increase in second harmonic signal for decreasing temperature, $\lambda_m^{2\omega}$ can be extracted very accurately in this regime, explaining the small errorbars on $\lambda_m^{2\omega}$ here. 
	}
	\label{fig:lambda_m}
\end{figure}

Fig.~\ref{fig:lambda_m} shows the magnon spin diffusion length $\lambda_m^{1\omega}$ ($\lambda_m^{2\omega}$) that we found from the first (second) harmonic signals, as a function of temperature. It can be seen that $\lambda_m^{1\omega}$ and $\lambda_m^{2\omega}$ approximately agree within the experimental error, which further supports our claim that there is no difference in the transport mechanism for electrically and thermally excited magnons. Furthermore, there is only a small change in $\lambda_m$ over the probed temperature range. Since $\lambda_m=v_{\rm th} \sqrt{\frac{2}{3} \tau \tau_{\rm mr}}$ \cite{Cornelissen2016a}, with $v_{\rm th}$ the magnon thermal velocity, $\tau$ the momentum relaxation time and $\tau_{\rm mr}$ the magnon spin relaxation time, we attribute this to the fact that while the relaxation times increase as $T$ decreases, this is compensated by the reduction in thermal velocity of the magnons. Our results differ significantly from the findings of Giles \emph{et al.}, who reported as a lower bound $\lambda_m=43$ $\upmu$m for at 23 K (compared to $\lambda_m(T=25)=6.7\pm0.6$ $\upmu$m that we find here). Note that we study a 210 nm thin YIG film on GGG substrate, whereas Giles \emph{et al.} studied a 0.5 mm thick YIG substrate. However, recent magnon spin transport studies in our group did not show significant variation in $\lambda_m$ for YIG film thicknesses up to 50 $\upmu$m at room temperature \cite{Shan}.

\begin{figure*}
	\includegraphics[width=17.5cm]{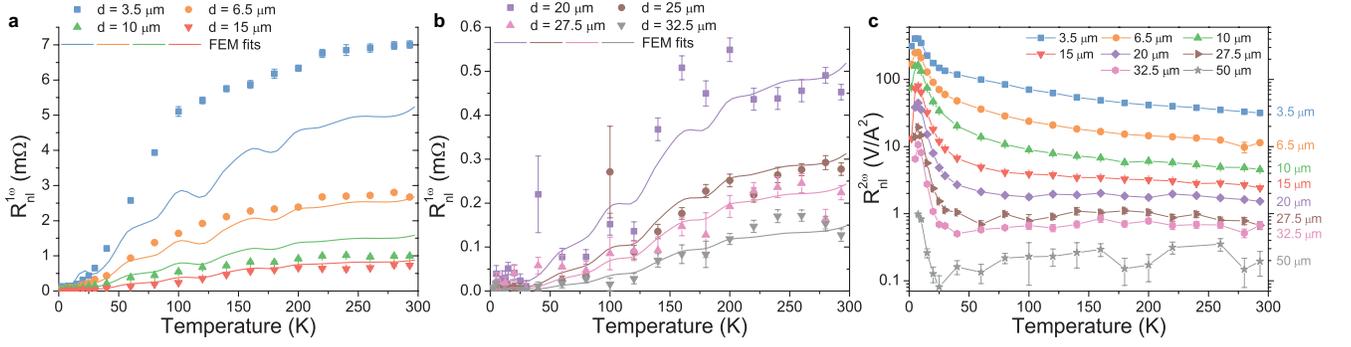}
	\caption{Replot of the data shown already partially in Fig.~\ref{fig:dist_dep}, now as a function of temperature. (a) and (b) Amplitudes of the first harmonic non-local resistance (symbols) for various injector-detector distances as a function of temperature. Solid lines show the results of the temperature dependent 2D FEM for every distance. In the FEM, $\sigma_m$ is used as the only free parameter to fit the data from all distances at each temperature. The drawn fit results (solid lines) are a guide to the eye, and the non-monotonous variation in these lines is very likely non-significant. (c) Amplitude of the second harmonic non-local resistance for various distances as a function of temperature, on a logarithmic scale. The signal peaks at $T\approx7$~K for all distances.
	}
	\label{fig:exp_model_comparison}
\end{figure*}

Our observed $\lambda_m(T)$ also differs from recent experiments which rely on the YIG thickness dependence of the local spin Seebeck effect (SSE) to determine the propagation length $\xi$ of thermally excited magnons \cite{Guo}. The authors of Ref.~\cite{Guo} found a scaling of $\xi\propto T^{-1}$, and this completely different temperature dependence might indicate that the local SSE is governed by a different length scale than its non-local counterpart which we study here.

Focussing on the electrical generation of magnons, we use the 2D finite element model which we developed in Ref.~\cite{Cornelissen2016a} to describe the first harmonic non-local resistance as a function of temperature. The model, which is based on the linear-response transport theory for the diffusive spin and heat transport of magnons, is described in detail in Ref.~\cite{Cornelissen2016a}. Magnon spin transport in the bulk of the YIG (in corresponding electrical units) is described by
\begin{align}
\frac{2e}{\hbar}\vec{j}_m &= -\sigma_m \vec{\nabla} \mu_m, \\
\nabla^2\mu_m &=\frac{\mu_m}{\lambda_m^2},
\end{align}
where $\vec{j}_m$ is the magnon spin current, $e$ is the electron charge, $\hbar$ the reduced Planck's constant, $\sigma_m$ the magnon spin conductivity, $\mu_m$ the magnon chemical potential and $\lambda_m$ the magnon spin diffusion length. Spin currents across the Pt\textbar{}YIG interface are given by ${j_m^{\rm int}=g_s\left(\mu_s-\mu_m\right)}$, where $\mu_s$ is the spin accumulation at the Pt side of the interface, and $\mu_m$ the magnon chemical potential on the YIG side. In linear response, magnon spin transport in our Pt\textbar{}YIG devices is thus governed by three parameters: $\lambda_m$, $\sigma_m$ and $g_s$. 

Using our model, we aim to find $\sigma_m(T)$ which is thus treated as the only fit parameter. $\lambda_m(T)$ is found from the distance dependence of the non-local signals directly as shown in Fig.~\ref{fig:lambda_m} and we use $g_s(T) = g_s(293)(T/293)^{3/2}$, where $g_s(293)$ is the value for $g_s$ at room temperature ($293$~K). We extracted $G_r=2.5\times10^{14}$ S/m$^2$ from spin Hall magnetoresistance (SMR) measurements \cite{Nakayama2013, Vlietstra2013a} of our devices at room temperature, from which we obtain $g_s(293)=1.5\times10^{13}$~S/m$^2$, comparable to what we found for our previous devices in Refs.~\cite{Cornelissen2015,Cornelissen2016a}. Furthermore, the spin accumulation generated at the interface of the Pt injector and the YIG was calculated using ${\mu_s=2\theta_{\rm SH}j_c\frac{\lambda_s}{\sigma_e}\tanh\left(\frac{t}{2\lambda_s}\right)}$ \cite{Flipse2014,PhysRevB.88.094410}, where $\theta_{\rm SH}$ is the spin Hall angle in Pt, $j_c$ is the charge current density in the injector, $\lambda_s$ is the spin diffusion length in Pt, $\sigma_e$ the Pt conductivity and $t$ the Pt thickness. $\sigma_e(T)$ is extracted independently from resistivity measurements on the injector strips and also used as input in the FEM. Finally, the non-local signal is found by calculating the average spin current density $\langle j_s \rangle$ in the detector, which is then converted to non-local resistance using $R_{\rm nl}=\theta_{\rm SH}L \langle j_s \rangle /(I\sigma_e)$.

Fig.~\ref{fig:exp_model_comparison} shows the measured first harmonic non-local resistance as a function of temperature, for various distances. The solid lines are the results of the fit of the FEM to the experimental data, with $\sigma_m$ as the only free parameter. Fits are performed to data for all distances at each measured temperature to extract $\sigma_m(T)$ \footnote{We again fit the $\log_{10}$ of the data to the $\log_{10}$ of the model outcome to obtain equal weighting for data from all distances.}. The agreement between model and experiment is reasonable as the model is generally less than a factor of 2 off, even for $d=3.5$~$\upmu$m where the largest discrepancy is observed. 

\begin{figure}
	\includegraphics[width=8.0cm]{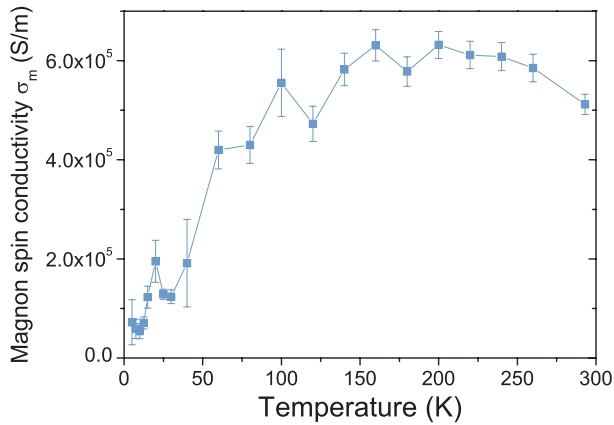}
	\caption{Magnon spin conductivity $\sigma_m(T)$ as a function of sample temperature, extracted from least squares fits of the 2D FEM to the experimental first harmonic data (Fig.~\ref{fig:exp_model_comparison}a and b). The FEM is used to fit the distance dependence of the signal at each temperature, with $\sigma_m(T)$ as the only free parameter. Errorbars indicate one standard error obtained from the fits. 
	}
	\label{fig:sigma_m}
\end{figure}

The resulting temperature dependence $\sigma_m(T)$ is shown in Fig.~\ref{fig:sigma_m}, where the errorbars indicate one standard error in $\sigma_m$ obtained from the fits. Note that the value we find for $\sigma_m$ at room temperature, $\sigma_m(293)=5.1\pm0.2\times10^{5}$~S/m, is consistent with $\sigma_m=5\times10^5$~S/m extracted previously from an independent set of data obtained from different devices (the majority of which were fabricated on samples cut from the same YIG\textbar{}GGG wafer) \cite{Cornelissen2016a}. 

Fig.~\ref{fig:exp_model_comparison}c shows that the second harmonic signal exhibits a maximum at $T\approx7$ K for all distances. Below $7$~K the signals decrease again, even changing sign for large distances ($d\geq20$~$\upmu$m) \footnote{Datapoints with negative amplitudes are not visible in Fig.~\ref{fig:exp_model_comparison}c due to the logarithmic scale of the plot. The sign changes occur in the region $2<T<5$~K, depending on the distance.}. This sign change is not well understood and calls for further investigations. In particular, a study of the non-local second harmonic signal temperature dependence as a function of YIG thickness may lead to more insight in the complicated generation mechanism for thermal magnon excitation, since recent experimental results show that the distance at which the sign change occurs (at room temperature) depends on the thickness of the YIG film, whereas $\lambda_m$ does not depend on film thickness \cite{Shan}.

The enhancement in second harmonic signal is at present not well understood. However, we do attribute it to an enhancement in thermal magnon generation at or close to the injector (rather than changes in the transport or detection of the magnons), since the only difference between first and second harmonic signal lies in the generation mechanism of the magnons. This could mean that the spin Seebeck coefficient in YIG is enhanced for decreasing temperature, however an extensive analysis is needed to draw further conclusions regarding the origin of this enhancement. Since we focussed here on the temperature dependence of the transport parameters involved, we leave this analysis for future work.

In conclusion, we report the temperature dependence of the magnon spin diffusion length and the magnon spin conductivity in YIG, which we extracted from non-local magnon spin transport measurements. We observe only a slight change in $\lambda_m$ with temperature, which we attribute to the fact that the increase in magnon relaxation time is compensated by the reduced thermal velocity of the magnons. The close agreement in $\lambda_m$ for electrically injected and thermally generated magnons confirms that the same (exchange) magnons are involved, and supports the description of the non-equilibrium transport in terms of a magnon chemical potential \cite{Cornelissen2016a}. For electrically generated magnons, we modeled the distance and temperature dependence of the non-local signal quantitatively using a 2D finite element model which was developed in earlier work. The model gives good agreement with the experimental observations over the whole temperature range and allowed us to find the temperature dependence of $\sigma_m$, which we find to decrease by roughly an order of magnitude from room temperature to $T=5$~K. For thermally generated magnons, we observe that the non-local signal increases with decreasing temperature and peaks around $T\approx7$~K. Additional experimental and theoretical studies are required to understand this enhancement. 

The authors would like to acknowledge H. M. de Roosz, J.G. Holstein, H. Adema and T.J. Schouten for technical assistance. This work is part of the research program of the Foundation for Fundamental Research on Matter (FOM) and supported by NanoLab NL, EU FP7 ICT Grant No. 612759 InSpin and the Zernike Institute for Advanced Materials.
\bibliography{main}
\end{document}